# Demonstration of Spatial Self Phase Modulation based photonic diode functionality in MoS$_2$/h-BN medium

Mahalingam Babu[1‡], Sudhakara Reddy Bongu[2‡], Pritam P Shetty[1], Eswaraiah Varrla[3],

G Ramachandra Reddy[4], Jayachandra Bingi[1, *]

[1] Bio-inspired Research and Development (BiRD) Laboratory, Photonic Devices and Sensors (PDS) Laboratory, Indian Institute of Information Technology Design and Manufacturing (IIITDM), Kancheepuram, Chennai 600127, India.

[2] Chair of Large Area Optoelectronics, University of Wuppertal, Rainer-Gruenter-Str. 21, 42119 Wuppertal, Germany.

[3] Nanosheets and Nanocomposites Laboratory, Department of Physics and Nanotechnology, SRM Institute of Science and Technology, Kattankulathur, Chengalpattu, Tamil Nadu-603203, India.

[4] Ravindra College of Engineering for Women, Kurnool, AP, India.

* Corresponding author, email-id: bingi@iiitdm.ac.in

‡ These authors contributed equally.

**Abstract**

Spatial self-phase modulation (SSPM) is the optical nonlinear process and is a result of spatially varying refractive index profile along the line of propagation in a medium. SSPM is proved to be a method to demonstrate different photonic functionalities. Transition metal dichalcogenides play a key role in 2D nanophononics due to their unique and fascinating properties. MoS$_2$ is the widely studied layered TMDs among all other 2D materials. This paper demonstrates such photonic functionality using thermally induced nonlinear optical response SSPM method, of MoS$_2$ nano bottles. Thermally induced nonlinear optical parameters have been estimated by utilizing the saturable absorption response of h- BN, the nonreciprocal light propagation has been achieved. The diode actions have also been demonstrated in liquid-solid and solid-solid devices with the help of passive elements.

## 1. Introduction

Spatial self-phase modulation (SSPM) in general the consequence of the optical Kerr effect observed when the medium is excited by a high intensity laser beam [1]. However, the changes in refractive index occur due to the cumulative thermal effect, which leads to the thermal induced SSPM. The phenomena is due to the creation of a thermal lens region by the laser and the self-interference of the laser beam due to the same region, where concentric rings are found in the far field region. The phenomena is demonstrated in a medium containing different nanostructures such as 2D sheets, tubes, spheres etc. and different biomaterials also used as samples[2]–[6]. The SSPM phenomena form the basis for different photonic applications such as optical bistability [1], vortex beam generation[7], all optical switch[3], optical limiting and diode applications[8][9][10] where modulation of concentric rings using the external stimuli is important.

On the other hand, the 2D materials such as graphene, $MoS_2$, h-BN etc. are being proven potential for photonic applications. Among the 2D materials $MoS_2$ in the form of 2D nanosheets, nanoparticles and as various nanostructures is potential for optoelectronic and photonic device applications. The $MoS_2$ 2D material (single layer as well as multi-layer forms) are demonstrated for micro lasers, exciton plasmon coupling, photoluminescence tailoring, optoelectronic memory, excitonic transistor and flexible photodetector due to the phenomena occurs in $MoS_2$ such as band renormalization, thermal excitons, exciton modes etc.

Further, SSPM phenomena is demonstrated in the $MoS_2$ liquid medium, the SSPM based all optical switching is also demonstrated with relatively large nonlinear dielectric susceptibility [11]. The $MoS_2/TiO_2$ nanocomposite [12] is used to demonstrate the quenched SSPM at lower wavelengths and enhanced SSPM at longer wavelengths. Same way usage of $MoS_2$ along with the other 2D materials such as h-BN may be useful to tune the wavelength dependent and direction dependent SSPM response.

This work proposes and demonstrates all optical diode phenomena using other 2D material and metallic coatings along with $MoS_2$. Also, liquid and solid state all optical diode using the SSPM is demonstrated.

## 2. Material Synthesis and experimental

### 2.1. $MoS_2$ and h-BN nanofluid preparation

We bought high quality $MoS_2$ and h-BN in powder forms from the Sigma- Aldrich (≥99.99% for purification) and used as received. The following process is used for preparing the $MoS_2$ and h-BN nanofluids. Initially, $MoS_2$ powder was grinded for 2.5 hours in a mortar. The stable fluid matrix was prepared by mixing 0.45 g of PVP with 14.55 g of ethanol-acetone binary solvents (7.27g of ethanol and 7.27g of acetonitrile). Further 3% of grinded $MoS_2$ powder is added to the as prepared fluid matrix. The mixture is subjected to vigorous agitation by ultrasonication for 3 hours. As received 2D crystal dispersion was centrifuged at 6000 rpm for 2 hours to separate the unexfoliated bulk MoS$_2$ sample. Further, the decanted dispersion was collected and again sonicated for an hour to make furthermore exfoliation and followed centrifugation at 6000 rpm for 5 min. In this process, we could achieve the highly uniform 2D MoS$_2$ nanofluid. It was found that that nanofluid was stable for several days. In a similar way, the h-BN nanofluid is prepared in the same amount of solvent (toluene) along with the 0.45g of polystyrene. Further processing is similar to that used for the MoS$_2$ nanofluid.

*2.2.Spatial self-phase modulation (SSPM) experiment*

The samples prepared in the aforementioned methodology are used to demonstrate the SSPM. The experimental setup used for the demonstration is as shown in Figure 1A, where the laser beam of 650 nm / 405 nm is passing through the neutral density filters and the convex lens of focal length 5 cm, the samples are kept 0.3cm after focal point. The achieved beam waist radiuses of the pump beams are 12.77 µm and 14.47 µm, respectively, corresponding to 405 nm and 650 nm. The lab made glass cuvette of thickness 1.5 mm used for the measurements. The converged laser beam is passed through the MoS$_2$ nanofluid. The separation between the lens and the MoS$_2$ sample is chosen in such a way to get the clear SSPM effect. Further, the transmitted signature beam from the sample is captured at the screen with a camera. The cuvette to screen separation is 140 cm. In this experiment, the ND filters are changed to tune the intensity at the sample.

Further for the SSPM measurements, the MoS$_2$ and h-BN are taken in the liquid form and loaded in to a single glass cell. The double cuvette system of 1.5mm path length each was used for the liquid based photonic diode measurements. Panel B demonstrates the formation of typical SSPM diffraction rings in MoS$_2$ nanofluid over a period of time. For the measurement of thermal induced nonlinear refraction coefficient, the SSPM diffraction rings from the sample were carried with the varying the pump fluence. In addition to these experiments the h-BN nanofluid was also replaced with ND filter and aluminum coating in the liquid-solid and solid-solid state devices, respectively. Soda lime glass slide is used as a substrate for solid state device. The general cleaning procedure for substrate is done as follows: Glass slide is sonicated in soap water for 5 minutes to remove oil and dust. The substrate is rinsed by using de-ionized water. Further, it is sonicated in Acetone and then in Methanol for 5 minutes each. The aluminum coating on the glass slide is done through DC sputtering with parameters Voltage: 402V, Current (I): 0.18 A, Pressure: 0.017 Pa, Deposition Time: 2 sec. On the other hand, the MoS$_2$ is coated on the other side of the same slide through a drop cast method (solvent evaporation). Further, it is allowed to dry naturally in ambient room temperature of 32°C for 30min. The experiments are performed with both liquid and solid-state devices at wavelengths of 405 and 650nm.

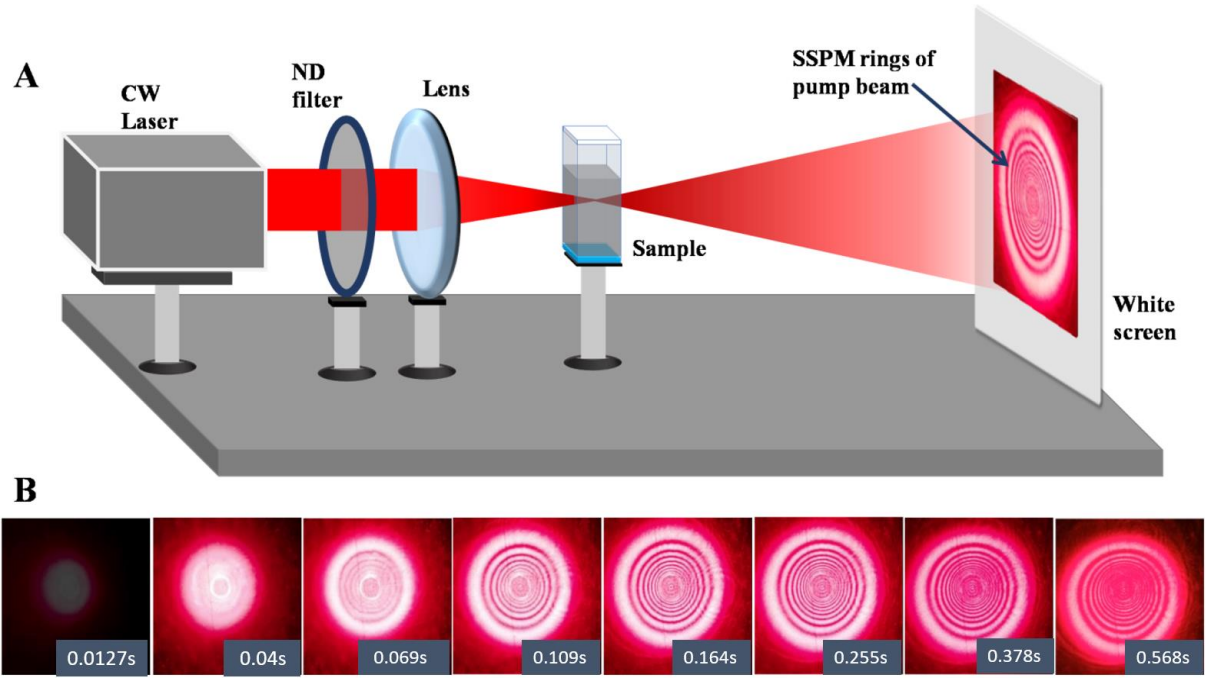

*Figure 1: Panel A, experimental layout of the spatial self-phase modulation setup. ND filter is neutral density filter for controlling pump power, CW laser is continuous wave laser with various wavelengths. Panel B, the transformation of SSPM rings in MoS$_2$ nano-fluid upon excitation with 650 nm CW laser (frames extracted from slow motion (1/8$^{th}$ of time) video in Supplementary).*

## 2.3. Characterization of nanomaterials

The MoS$_2$ and h-BN nano-fluids were characterized by the field-emission scanning electron microscopy (FE-SEM), Raman spectrometer and UV-Visible absorption spectroscopy. Panel A and B shows the photographic images of the as prepared nano-fluids of MoS$_2$ and h-BN, respectively. FE-SEM micrographs (Figure 2C and D) confirmed that MoS$_2$ nanofluid has a uniform bottle shape particle, has an average length of 1.85 µm, and a diameter of 186 nm. The Raman spectrum of MoS$_2$ (Figure 2E) shows the two prominent modes: an in-plane ($E^1_{2g}$) located at 381.79 cm$^{-1}$ and an-out of plane ($A_{1g}$) at 407.287 cm$^{-1}$. The $E^1_{2g}$ mode occurs as a result of two S atoms and one Mo atom vibrating opposite to each other while the $A_{1g}$ mode is due to vibrations between S atoms in the opposite direction. The interpeak frequency difference is 25.497 cm$^{-1}$ indicating MoS$_2$ nanoflakes are having more than six layers[13][14]. From the Figure 1F the Raman spectra of h-BN revealed $E_{2g}$ mode at 1323.86 cm$^{-1}$ which is due the in-plane B$_3$N$_3$ vibrations. The broadness of peak can be attributed to the strain and stretching of few layered h-BN flakes[15][16]. Strain can be induced into the flakes as a result of synthesis method, i.e., grinding and exfoliation in this research. Panel G gives the absorption spectrum $MoS_2$ nanofluid. Two sharp bands observed at 609 nm and 670 nm are the characteristic excitation bands A and B of the $MoS_2$ sample. In addition, a broad band appears at 434nm assigned to be C excitation peak. The excitations for the SSPM measurements were done at near to the C and B excitation bands. Panel I show bandgap structure, the $MoS_2$ and h-BN structures are synthesized after exfoliating from their bulk counterparts. $MoS_2$ is a semiconductor in the 2H phase having a bandgap of 1.5 eV with the number of layers less than ten whereas exfoliated h-BN is an insulator with a bandgap of 5.6 eV. These results are in line with the reported literature. The thickness effect is dominant in $MoS_2$ compared with h-BN due to differences in their bonding characteristics and interlayer coupling.

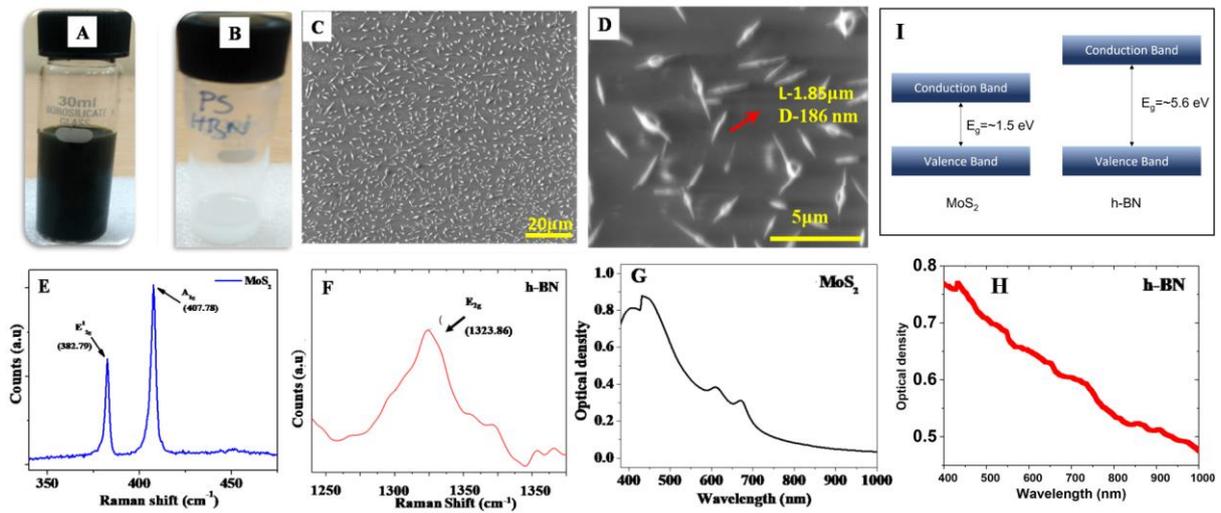

*Figure 2: Panel A and B show the photographic images of the stable nanofluid solutions of $MoS_2$ and h-BN, respectively. Electron microscopic images of $MoS_2$ (panel C and D). Panel E and F corresponding to the Raman spectra of MoS2 and h-BN, respectively. Absorption spectrum of $MoS_2$ nanofluid and h-BN solution (panel G and H). Bandgap structure of $MoS_2$ and h-BN used in current work (Panel I).*

### 3. Results and Discussion

Figure 3a shows the SSPM rings patterns of $MoS_2$ nanofluid with various pump powers upon pumping with 650 nm and 405 nm. Once the pump beam with sufficient intensity passes through the $MoS_2$ sample, the transmitted beam expands and forms the ring patterns within milliseconds. Further, the size of the ring numbers reaches to a maximum in few seconds and forms the stable pattern (Figure 1B). The SSPM rings patterns were recorded at the screen by varying the pump power. It is clear from the visual inspection that the number of rings is changing w.r.t the incident power. Hence, the $MoS_2$ nanofluid is exhibiting the SSPM in the two extremes of the visible spectra. Compared to the previous reports, the current measurement exhibits the stable and nondestructive SSPM patterns in MoS$_2$ nanofluid. It can be due to the high stability and unique shape of the MoS$_2$ nano bottles [4], [17], [18]. Wei et al., have found the effect of gravitational force on the symmetry of SSPM pattern in 1D carbon nanotubes [17]. Yun et al, have reported the distortion of SSPM pattern in 2D MXene nano sheets in the time scale of 1 sec, which is attributed to the asymmetric thermal convection near the focal point[18], [19]. In the present case the SSPM pattern in MoS$_2$ nano bottles slightly distracted from the symmetric shape at a longer time scale suggesting the less divergence in the density of MoS$_2$ nanobottles during the SSPM. This can be due to the highly stable and unique shape of the MoS$_2$ nanostructures. SSPM is the consequence of the optical Kerr effect where the intense laser pulses alters the refractive index of the medium. However, in current case the modifications in the refractive index occur due to thermal induced nonlinear refraction. The large amount of the pump energy absorbed by the photonic medium delivers the local heat. Such locale heat results in the cumulative thermal effects inside a sample that alters the refractive index of the medium. The alterations in refractive index of the medium due to thermal effect act as a lens and known as thermal lens. Most of the solvents posses' the negative thermal induced nonlinear response resulting in negative lensing effect. The parameters characterize the thermal lens, thermal rise and relaxation dynamics of the interacting medium typically in the order of ns and ms, respectively. Under the CW pumping a stationary thermal lens will form inside a medium and responsible for the observed SSPM ring patterns in the MoS$_2$ nanofluid.

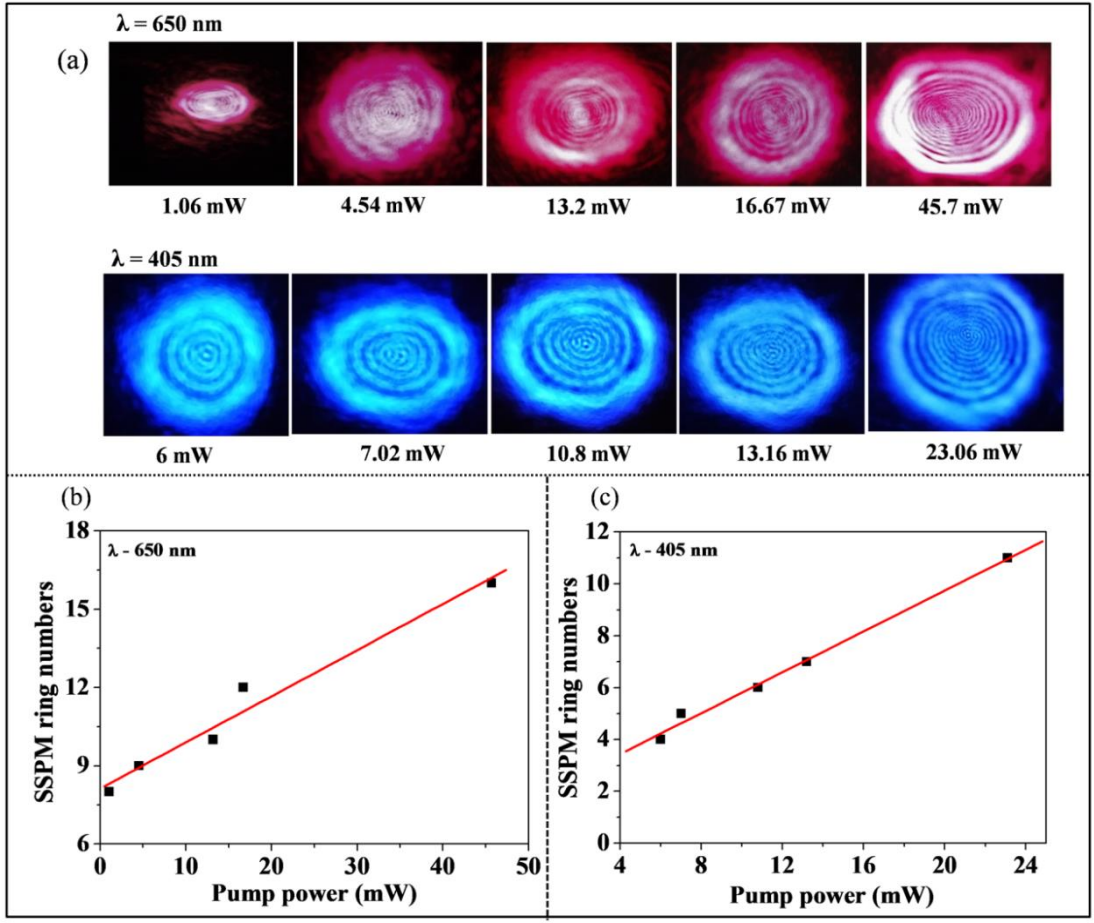

*Figure 3: Panel (a), SSPM pattern in MoS₂ nanofluid on various pump powers at λ= 650 nm and at λ= 405 nm. Panel (b) and (c): dependence of the SSPM ring numbers on the pump powers at λ= 650 nm and at λ= 405 nm, respectively.*

Figure 3b and 3c gives the number of SSPM rings as function of pump intensity for both wavelengths 650 nm and 405 nm, respectively. The refractive indices are calculated based in the number of SSPM rings using the following formula and theory[20][21].

$$n_2 = \frac{\lambda}{2n_0 L_{eff}} \cdot \frac{N}{I}$$

Here, $\lambda$ is the pump wavelength, $n_0$ is the linear refractive index, $L_{eff}$ (= $1 - \exp(-\alpha_0 L)/\alpha_0$) is the effective path length, N is the number of SSPM rings and I is the pump irradiance. For input pump power $P$, we can calculate total irradiance $I$ falling on the sample by $I = 2P/\pi\omega^2$. The values of N/I determined form the linear fit of Figures (b) and (c) for both pump wavelengths. The deduced thermal induced nonlinear refractive indices of MoS₂ at 650 nm and at 450 nm are $1.069 \times 10^{-11}$ m²/W and, $1.387 \times 10^{-11}$ m²/W respectively. The large value of nonlinear refractive index at lower wavelengths is attributed to the stronger absorption at the C excitation peak of MoS₂. The higher absorption of pump light by nonradioactive photonic medium leads to larger accumulation of local heat inside a sample that result in formation of stronger thermal lens.

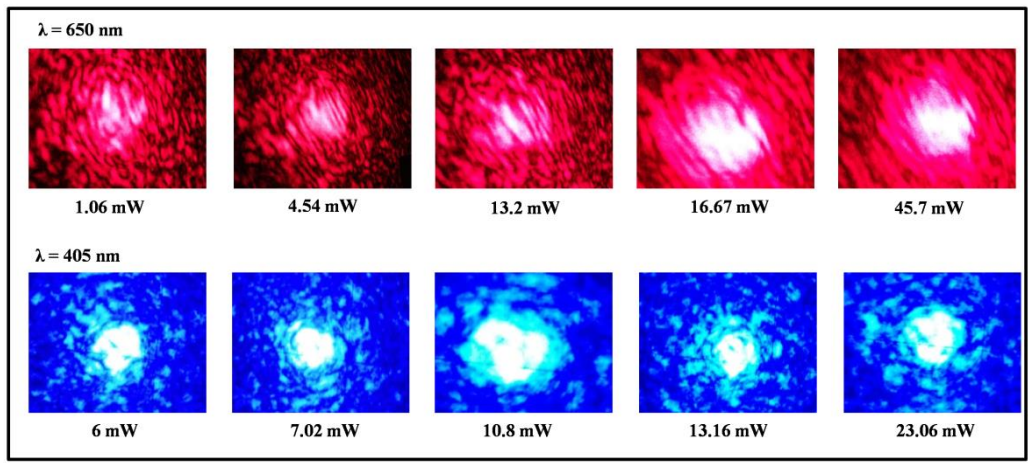

*Figure 4: The optical signature from the h-BN sample in the same condition of MoS$_2$ SSPM experiment*

Further, the same experiment has been performed on the h-BN nanofluid with identical experimental conditions. It is observed that the SSPM effect is not occurring in the h-BN sample for the defined conditions. The high scattering beam with speckle formation is observed on the screen as shown in the Figure 4. This clearly indicates the defined power and intensity conditions for MoS$_2$ nanofluid are not enough to trigger the SSPM in the medium. Hence, the h-BN nanofluid can act as a reverse saturable absorber in a device along with MoS$_2$. To demonstrate the optical diode phenomenon, it is important to load the layer that will not respond w.r.t SSPM.

### 3.1. MoS$_2$ +h-BN for optical diode

The double cuvette sample holder is prepared to accommodate both the MoS$_2$ and h-BN nanofluids. The experimental configuration is the same as used earlier. Here, the experiment is performed at different wavelengths (650 nm and 405 nm) and in two sample configurations, forward and backward directions as shown in Figure 5a. In forward (MoS$_2$ → h-BN) direction the pump beam initially passes through the MoS$_2$ sample while it passes through the h-BN in backward (h-BN → MoS$_2$) direction. The forward configuration clearly gave the SSPM patterns both at 650 nm and the 405 nm. Whereas the reverse configuration resulted in the pure scattering speckle patterns due to non-SSPM response of the h-BN. The power dependency of the SSPM is still valid with an increasing number of rings with the pump power. Panel (b) and (c) gives the number of SSPM rings as function of pump power in both forward and back ward directions at 650 nm and at 405 nm, respectively. The nonreciprocal light transmittance of the device can be observed at both wavelengths which is the signature of a photonic diode. Compared to the pumping at 650 nm, 405 nm show the large effect as observed in individual samples due to aforementioned mechanism.

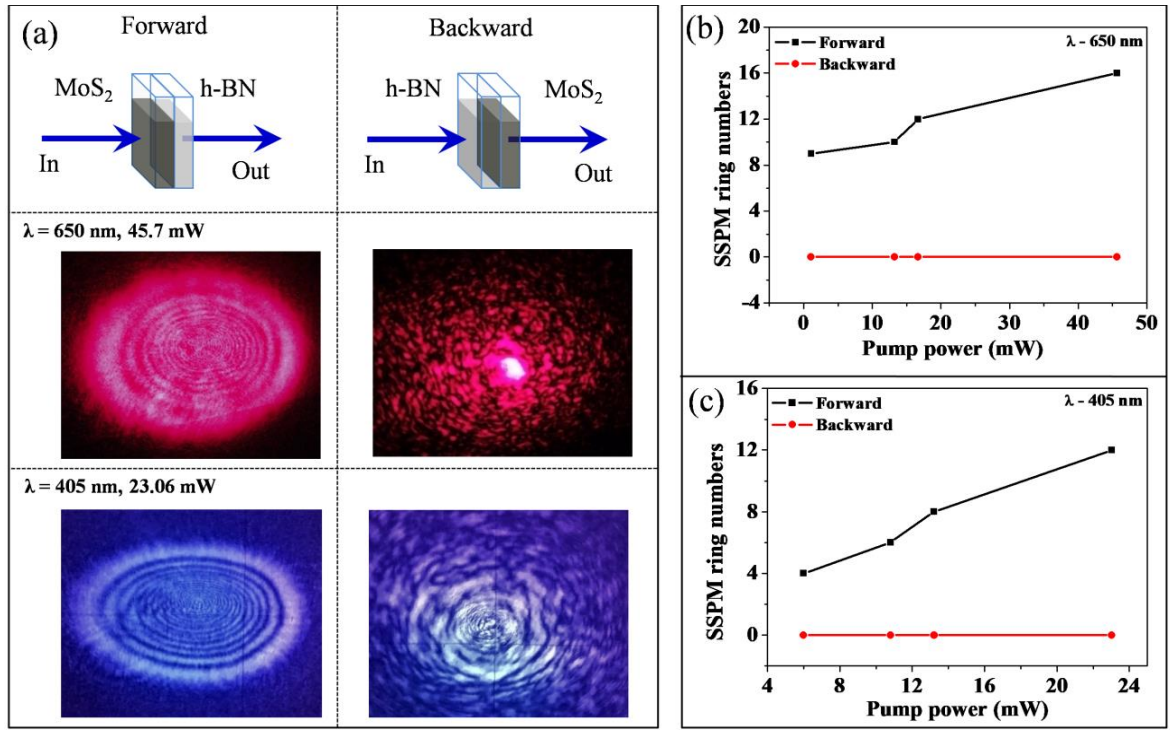

*Figure 5: (a) Sample configuration for the liquid based photonic diode demonstration and SSPM rings in forward and backward directions for the both wavelengths at their highest power. The results of SSPM from the proposed photonic diode at 650 nm (b) and 405 nm (c), respectively.*

The fluctuation in the number of rings at low power incidence were observed may be attributed to the disturbance (scattering) created by the h-BN medium in the forward propagation of SSPM pattern. Further, the nanofluid solvent affects the thermal fluctuations within the medium with respect to time, which may be another reason for ring fluctuations. To subside the impact of the solvent and to minimize the scattering introduced by the h-BN medium, the device is upgraded to the liquid-solid and solid-solid state by replacing the h-BN medium with ND filter and aluminum coating which work as a semitransparent layer, respectively.

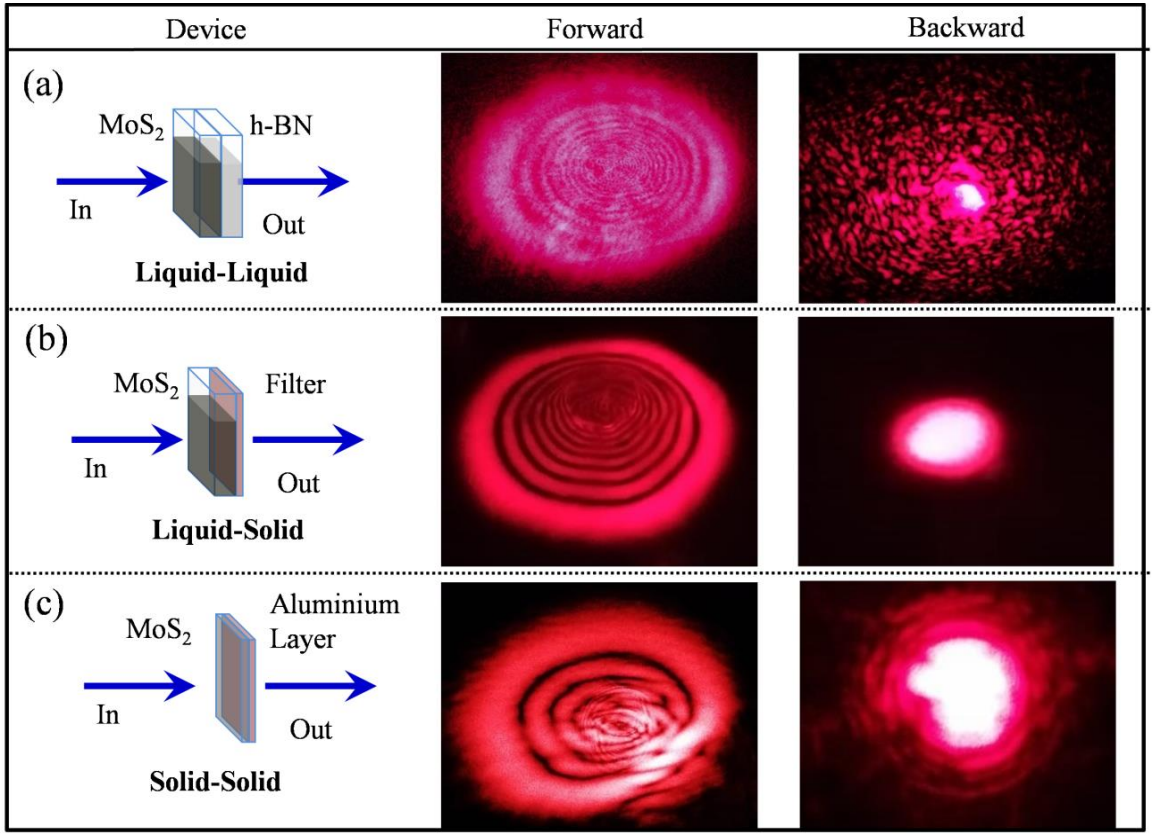

*Figure 6: Various photonic device fabrications with MoS$_2$ nanofluid and SSPM pattern in the forward and backward directions: (a) a liquid-liquid device; (b) a liquid-solid device; and (c) a solid-solid device.*

The solid-state device is prepared with MoS$_2$ and Aluminum coating on either side of the glass substrate. This experimental setup includes the pinhole to avoid the burning effects and divergent scattering effects from the laser. As the beam diameter is reduced due to pinhole the device is moved towards the focal region to observe the SSPM effect in the forward direction. In the same manner the device is turned where the aluminum layer is exposed to the focal region in the reverse configuration. As expected, the aluminum in the reverse configuration reduces the incident intensity besides its inability to give SSPM pattern, which lead to the concentrated power output on the screen. As shown in Figure 6 (b)and (c) the liquid-solid and solid-solid devices results in the well-defined SSPM patterns in the forward configuration and stable with concentrated power dot output in reverse configuration. These devices clearly avoid the scattering and solvent driven thermal effects in the output. Compared to the liquid-liquid and liquid-solid devices, solid-solid device plays a crucial role in fabrication solid state photonic device which avoids the usage of liquid samples. It is found that the spatial refractive index profile imprinted when experiment is conducted with high beam power of 45mW. In the Liquid-liquid, liquid-solid, and solid-solid devices the fundamental SSPM mechanism is impacted by, the pathlength, scattering due to particle motion, thermal diffusion and dissipation.

## 4. Conclusion

This research uses the $MoS_2$ nano-fluid with nano bottles as medium to demonstrate the SSPM effect. Thermal induced nonlinear refractive indices are calculated at near C and B excitation bands of $MoS_2$. The $MoS_2$+h-BN structure in the liquid form is used to demonstrate the photonic diode phenomenon. The observed thermal fluctuations and the scattering effects are avoided by liquid-solid and solid-solid state devices with ND filter and $MoS_2$+Al coating combination, respectively, where the clear and stable photonic diode functionality is demonstrated. The demonstration in solid-solid device finds the way to avoiding the liquid samples. This work has the potential to show the $MoS_2$ as a solid-state photonic device.

## 5. Author contributions

**Mahalingam Babu**: Nanofluid synthesis, Validation, Investigation, **S R Bongu**: Conceptualization, Methodology, Data analysis, Writing - Review & Editing, **Pritam P Shetty**: Investigation, Visualization, Writing - Original Draft, **Eswaraiah Varrla**-Characterization, Writing - Review & Editing, **G Ramachachandra Reddy**-Writing - Review & Editing, **Jayachandra Bingi**: Conceptualization, Methodology, Writing - Review & Editing, Supervision.

## 6. Declaration of Competing Interest

The authors declare that they have no known competing financial interests or personal relationships that could have appeared to influence the work reported in this paper.

## 7. Acknowledgements

Authors acknowledge the funding support from DST India under INT/RUS/RFBR/P-262. Authors acknowledge Dr. PV Karthik Yadav and Dr. Y Ashok Kumar Reddy for providing support in sputter coating.